\documentclass[aps,floats,floatfix,showpacs,preprintnumbers,amssymb,prd,twocolumn,superscriptaddress,nofootinbib,nolongbibliography,reprint]{revtex4-1}

\usepackage[T1]{fontenc}
\usepackage[utf8]{inputenc}               
\usepackage{hyperref}
\hypersetup{
    colorlinks=true,
    linkcolor=blue,
    citecolor=blue,
    urlcolor=blue
}

\usepackage{amssymb,amsmath,verbatim,mathtools,needspace,enumitem,etoolbox,graphicx,physics,microtype,afterpage,xspace,tabularx,lmodern,multirow,boldline, makecell, booktabs}

\usepackage{appendix}
\usepackage{tensor}
\usepackage{array}
\usepackage{siunitx}
\usepackage[normalem]{ulem}
\usepackage[dvipsnames, usenames]{xcolor}
\usepackage{xr-hyper}
\usepackage[font=small,labelfont=bf]{caption}
\definecolor{linkcolor}{rgb}{0.0,0.3,0.5}
\usepackage[all]{hypcap}
\usepackage[T1]{fontenc}
\usepackage[utf8]{inputenc}
\usepackage[usenames,dvipsnames]{xcolor}
\definecolor{cornellGreen}{HTML}{6EB43F}
\definecolor{cornellRed}{HTML}{B31B1B}
\hypersetup{colorlinks=true,citecolor=cornellRed,linkcolor=cornellRed,urlcolor=cornellRed}
\usepackage{multirow}

\definecolor{romared}{RGB}{142,0,28}

\def\bb#1\ee{\begin{equation}\begin{split}#1\end{split}\end{equation}}

\def\({\left(}
\def\){\right)}
\def\[{\left[}
\def\]{\right]}

\newcommand{\bd}{{\dot{\beta}}} 
    
\newcommand{\pa}{\partial} 
\def\a{\alpha}
\def\L{\mathcal{L}}

\def\pa{\partial}

\def\be{\begin{equation}}
\def\ee{\end{equation}}
\newcommand{\beq}{\begin{eqnarray}}
\newcommand{\eeq}{\end{eqnarray}}

\usepackage{makecell}
\usepackage{soul}

\newcolumntype{Y}{>{\centering\arraybackslash}X}

\usepackage{graphicx}
\usepackage{dcolumn}
\usepackage{bm}
\usepackage{hyperref}
\usepackage{color}

\usepackage{times} 

\usepackage{graphicx} 
\usepackage{booktabs} 
\usepackage[font=small,labelfont=bf]{caption} 
\usepackage{amsfonts, amsmath, amsthm, amssymb} 

\usepackage{wrapfig} 
 \definecolor{mypurple}{RGB}{130, 0, 130} 

\begin{document}
\title{The Triple $T\bar{T}$-Like Flow in Quantum Field Theories: Irrelevant, Marginal, and Relevant}
\author{H. Babaei-Aghbolagh}
\email{hosseinbabaei@nbu.edu.cn}
\affiliation{Institute of Fundamental Physics and Quantum Technology, \& School of Physical Science and Technology, Ningbo University, Ningbo, Zhejiang 315211, China}
\author{Bin Chen}
\email{chenbin1@nbu.edu.cn}
\affiliation{Institute of Fundamental Physics and Quantum Technology, \& School of Physical Science and Technology, Ningbo University, Ningbo, Zhejiang 315211, China}
\affiliation{School of Physics, Peking University, \& Center for High Energy Physics,  No.5 Yiheyuan Rd, Beijing
100871, P.R. China}
\author{Song He }
\email{ hesong@nbu.edu.cn}
\affiliation{Institute of Fundamental Physics and Quantum Technology, \& School of Physical Science and Technology, Ningbo University, Ningbo, Zhejiang 315211, China}
\affiliation{Max Planck Institute for Gravitational Physics (Albert Einstein Institute), Am M\"uhlenberg 1, 14476 Golm, Germany}
\author{Jue Hou }
\email{jue.1.hou@kcl.ac.uk}
\affiliation{School of Physics \& Shing-Tung Yau Center, Southeast University, Nanjing 211189, P. R. China}
\affiliation{Department of Mathematics, King’s College London, The Strand, London WC2R 2LS, United Kingdom}

\date{\today}
\begin{abstract}
We introduce a one-parameter root-$T\bar T$-like flow,
$ \partial_\lambda \mathcal{L}=\mathcal{R}_\lambda^{1/\alpha}$,
which organizes stress-tensor deformations into irrelevant, marginal,
and relevant branches. Within duality-invariant electrodynamics in four dimensions, and equivalently within two-dimensional integrable sigma models, the flow admits a closed-form solution controlled by an auxiliary equation. The marginal point $\alpha=1$ reproduces the root-$T\bar T$ / ModMax branch, while $\alpha<1$ gives irrelevant deformations distinct
from the standard Born-Infeld $T\bar T$ flow. For $\alpha>1$, the same construction yields explicit relevant $T\bar T$-like Lagrangians. These results suggest that root-$T\bar T$ flows provide a common organizing principle for duality-invariant and integrable deformations.

\end{abstract}

\maketitle

\section{Introduction}
In modern physics, \textit{integrability} and \textit{electromagnetic duality} play a profound role, especially in the investigation of nonperturbative aspects of theories.  Integrability, typically characterized by an infinite set of conserved charges~\cite{Noether_1971} or a flat Lax pair~\cite{Maillet:1985ek,Maillet:1985ec,Chen:2021aid}, imposes a specific differential equation on $\mathcal{L}$ for two-dimensional systems. And electromagnetic duality, which specifically persists in many four-dimensional nonlinear electromagnetic theories, requires the duality invariant  condition~\cite{bialynicki1983nonlinear,Gaillard:1981rj,Gibbons:1995ap,Gibbons:1995cv,Avetisyan:2021heg}. Surprisingly, 
as shown in Ref.~\cite{Babaei-Aghbolagh:2025uoz}, there exists a unification formalism between four-dimensional duality invariant electromagnetic theory and two-dimensional integrable theories. This is achieved by unifying the duality invariant electromagnetic condition~\cite{bialynicki1983nonlinear,Gaillard:1981rj,Gibbons:1995ap,Gibbons:1995cv,Avetisyan:2021heg} and the integrability condition~\cite{Borsato:2022tmu,Ferko:2023wyi,Babaei-Aghbolagh:2025hlm,Sakamoto:2025hwi,Fukushima:2025tlj,Baglioni:2025tsc} into an isomorphic partial differential equation (PDE) in two independent variables. The specific symmetry preserved by the flow—integrability or duality—is determined solely by the identification of the two independent variables.

Both the Principal Chiral Model~\cite{Gell-Mann:1960mvl,Gross:1973id,Politzer:1973fx} (a seed for integrable sigma models) and Maxwell theory (a seed for nonlinear electrodynamics) generate families of $T\bar{T}$ deformed theories via a coupling parameter $(\lambda)$~\cite{Smirnov:2016lqw,Cavaglia:2016oda,Conti:2018jho,Babaei-Aghbolagh:2020kjg,Ferko:2022iru,Ferko:2024yua,Babaei-Aghbolagh:2025cni}, see~\cite{Jiang:2019epa,He:2025ppz} for a review. These deformations are governed by ``flow equations,'' which fall into three classes: \textit{irrelevant }, \textit{marginal}, and \textit{relevant } flows. Although explicit theories are known for the first two classes—exemplified by Born-Infeld theory~\cite{Born:1934gh} as an irrelevant  $T\bar{T}$-like flow~\cite{Conti:2018jho} and ModMax theory~\cite{Bandos:2020jsw} as a marginal root-$\text{T}\bar{\text{T}}$-like flow~\cite{Babaei-Aghbolagh:2022uij}—a Lagrangian model for the \textit{relevant } flow has remained conspicuously absent.

In this Letter, we demonstrate that all three-flow classes can be universally derived and classified through a single, master equation constructed from composite operators of the energy-momentum tensor by ``root-$T\bar{T}$ triple operator'':
\begin{equation}\label{geometries}
	\frac{\partial \L}{\partial \lambda}= \(\mathcal{R} _\lambda\)^{\frac{1}{\alpha}} \equiv \left\lbrace\begin{matrix}
		& \mbox{\it irrelevant}&\ \ {\rm for\ } \alpha <  1,\cr
		& \mbox{\it marginal}&\ \ {\rm for\ } \alpha =  1,\cr
		&\mbox{\it relevant}&\ \  {\rm for\ } \alpha > 1.
	\end{matrix}\right. 
\end{equation}
where $\alpha$ is constant and  $\mathcal{R}_\lambda=  \tfrac{1}{\sqrt{d}}\sqrt{ {T}_{\mu \nu}\,{T}^{\mu \nu}-\frac{1}{d} T^{\mu}{}_\mu T^{\nu}{}_\nu}$  is the marginal root-$T\bar{T}$ operator\cite{Babaei-Aghbolagh:2022uij,Babaei-Aghbolagh:2022leo,Ferko:2022cix,Conti:2022egv}. This classification is independent of spacetime dimension and field content.

The principal finding of this letter is a triple $T\bar{T}$-like deformed theory that encompasses all three classes of flow equations. This is achieved through a novel formalism, parameterized by $\alpha$, which satisfies what we term the ``root-$T\bar{T}$ triple flow equation''~\eqref{geometries}. This unique closed-form framework provides a simultaneous description of both four-dimensional electromagnetic theories and two-dimensional integrable theories. It achieves this by employing an auxiliary equation and drawing conceptual parallels to the approaches of Courant-Hilbert~\cite{10.1115/1.3630089} and the Russo-Townsend formalism~\cite{Russo:2025fuc}. Within our construction, this auxiliary equation emerges directly from the root-$T\bar{T}$ triple flow equation itself. A generalization of $T\bar{T}$ in two dimensions, known as $T\bar{T} + \Lambda_2$ with $\Lambda_2 \equiv \Lambda_2(\lambda)$ a coupling-dependent cosmological constant, was introduced in Ref.~\cite{Gorbenko:2018oov}. We extend this framework to include an arbitrary coupling-dependent cosmological term $\mathcal{R}_\lambda + f(\lambda)$ and explore higher-order deformations.
References~\cite{Kuzenko:2000tg,Kuzenko:2000uh,Kuzenko:2021qcx, Kuzenko:2023ebe, Kuzenko:2026vir} developed a general formalism for duality rotations of bosonic conformal spin-$s$ gauge fields ($s \geq 2$) in a conformally flat four-dimensional spacetime. For $s = 1$, this formalism reduces to $\mathsf{U}(1)$ duality-invariant nonlinear electrodynamics. For each $s \geq 2$, families of such models exist, including a higher-spin generalization of ModMax~\cite{Kuzenko:2021qcx, Kuzenko:2021cvx}. Every $\mathsf{U}(1)$ duality-invariant model is self-dual under a Legendre transformation, and Kuzenko has shown that each admits a higher-spin extension~\cite{Kuzenko:2026vir}. Our results show that fermions and higher‑spin fields introduce no new obstructions: the triple‑flow formalism provides a unified and universal classification of duality‑invariant theories.

\section{ Unification formalism }
The self-duality condition for nonlinear electrodynamics, when expressed in its differential form, constitutes a fundamental constraint on the Lagrangian density $\mathcal{L}$. This constraint takes the form of a nonlinear (PDE) governing the functional dependence of $\mathcal{L}$ on the two fundamental Lorentz-invariant scalar variables.

Electromagnetic duality in four-dimensional nonlinear electrodynamics manifests as a self-duality constraint on
$\mathcal{L}(F_{\mu\nu})$, with $F_{\mu\nu} = \partial_\mu A_\nu -\partial_\nu A_\mu $~\cite{bialynicki1983nonlinear,Gibbons:1995ap,Gibbons:1995cv,Gaillard:1981rj,Avetisyan:2021heg}:
\begin{equation}
\label{self_dualityggff}
F_{\mu\nu}\, \widetilde{F}^{\mu\nu} - G_{\mu\nu} \widetilde{G}^{\mu\nu} = 0\,,
\end{equation}
where $G_{\mu\nu} = -2 \partial\mathcal{L}/\partial F^{\mu\nu}$, and $\widetilde{G}^{\mu\nu}$ is the Hodge dual of $G^{\mu\nu}$. This self-duality condition reduces to the following PDE in $S = -\frac{1}{4} F_{\mu\nu} F^{\mu\nu}$ and $P = -\frac{1}{4} F_{\mu\nu} \widetilde{F}^{\mu\nu}$~\cite{bialynicki1983nonlinear,Sorokin:2021tge}:
\begin{equation}
\label{eq:lagrange_self_duality}
\left( \partial_S \mathcal{L} \right)^2 - 2 \frac{S}{P} \left( \partial_S \mathcal{L} \right) \left( \partial_P \mathcal{L} \right) - \left( \partial_P \mathcal{L} \right)^2 = 1\,,
\end{equation}
where the derivatives $\partial_S \mathcal{L}$ and $\partial_P \mathcal{L}$ denote the partial derivatives of the Lagrangian with respect to the electromagnetic invariant variables $S$ and $P$.
A Lagrangian $\mathcal{L}(S, P)$ satisfying Equation~\eqref{eq:lagrange_self_duality} describes a theory whose equations of motion are invariant under the duality transformation rotations in the space of electric and magnetic fields, a property characteristic of self-dual theories such as the well-known Born-Infeld electrodynamics. 
A reformulation of the duality-invariant condition in~\eqref{eq:lagrange_self_duality} can be obtained by the following variable change:
\begin{equation}\label{UV}
	U=\frac{1}{2} \bigg( \sqrt{P^2 + S^2}-S \bigg)
	,\quad
	V=\frac{1}{2} \bigg( \sqrt{P^2 + S^2}+S\bigg),
\end{equation}
where $U $ and $V $ are non-negative parity-even scalars.
The condition that $\mathcal L$ must satisfy to exhibit self-duality assumes the following simple form when $\mathcal L$ is expressed as a function of  
$(U,V) $:
\begin{eqnarray}\label{sdc1}
{\cal L}_U {\cal L}_V=-1.
\end{eqnarray}

Remarkably, the distinct conditions for duality invariance and integrability share an identical mathematical structure when expressed in terms of two appropriate theory-specific variables. For integrable sigma models, these variables—denoted $(P_1, P_2)$—are defined in light-cone coordinates via  currents $j = g^{-1}dg$ with components $j_\pm = j_0 \pm j_1$, producing scalar invariants
\begin{equation}
P_1 = -\text{tr}[j_+ j_-], \quad
P_2 = \tfrac12\left( \text{tr}[j_+ j_+]\,\text{tr}[j_- j_-] + (\text{tr}[j_+ j_-])^2 \right).\nonumber
\end{equation}
Drawing inspiration from duality-invariant electromagnetic theory, two new variables are defined as follows~\cite{Babaei-Aghbolagh:2025uoz}:
\begin{equation}\label{uuvv}
u = \tfrac14\big( \sqrt{2P_2 - P_1^2} - P_1 \big), \, v = \tfrac14\big( \sqrt{2P_2 - P_1^2}+ P_1 \big),
\end{equation}
In terms of these two variables, the integrability condition leading to the Lax flatness condition reduces to a condition on the Lagrangian as follows:
\begin{eqnarray}\label{sdcuv}
{\cal L}_u {\cal L}_v=-1.
\end{eqnarray}
Two well-known approaches to solving the differential equations in Eqs.~\eqref{sdc1} and~\eqref{sdcuv} are the Courant-Hilbert method and the Russo-Townsend auxiliary-field formalism. These approaches provide a general Lagrangian that requires solving an auxiliary equation, and a one-to-one correspondence can be established between their solutions.

For each pair $(U, V)$ in the positive quadrant in duality-invariant electromagnetic theories, we define $\mathcal{L}(U, V)$. Subject to condition $\mathcal{L}(0,V)=\ell(V)$, the Lagrangian is given by the characteristic ansatz:
\begin{equation}\label{Soul}
\mathcal{L} = \ell(\tau) - \frac{2U}{\ell'(\tau)},
\qquad
\tau = V + \frac{U}{\ell'(\tau)^2},
\end{equation}
where $\ell(\tau)$ is an arbitrary generating function and $\ell'(\tau) = \partial \ell / \partial \tau$.

A general  new auxiliary-field  solution for duality-invariant electromagnetic theories is encoded in the Lagrangian:
\begin{equation}\label{LAFF}
\mathcal{L}(U,V) = -\frac{U}{y} + y V - \Omega(y),
\end{equation}
where $\Omega(y)$ is a master potential determined by dualization constraints that $y = e^{\varphi}$ and $\varphi$ is an auxiliary field depending on $U$ and $V$. The equation of motion for $\varphi$ is given by $\partial_\varphi \mathcal{L} = 0$,  which yields the auxiliary equation:
\begin{equation}
\Omega'(y) = \frac{U}{y^2} + V.
\end{equation}
The new auxiliary-field approach is fully equivalent to the original formulation in terms of $\ell(\tau)$, related by a simple change of variables. This relation is given by the Legendre transformation:
\begin{equation}\label{eq:legendre}
\ell(\tau) = \tau y - \Omega(y), \qquad \tau = \Omega'(y), \qquad y = \ell'(\tau)\,.
\end{equation}

Applying the unification formalism to both approaches, we adopt the notation of duality-invariant electromagnetic theories with Lagrangian $\mathcal{L}(U, V)$, which can be translated into two-dimensional integrable theories via the dictionary $(U, V) \to (u, v)$. Under this mapping, we obtain $\mathcal{L}(U,V) \to \mathcal{L}(u, v)$.

\section{ A Triple Solution }\label{unif}
Every interacting theory parameterized by a coupling $\lambda$ can be viewed as a deformation of a seed theory. These deformations fall into three distinct classes — irrelevant, marginal, and relevant — each characterized by its flow equation.

We address the dual conditions of duality invariance and integrability, given by Eqs.~\eqref{sdc1} and~\eqref{sdcuv}, by starting from a seed theory. This seed theory is Maxwell's theory~($ \L_{Max}=V-U=S $), in the electromagnetic case, and a principal chiral model,~($ \L_{PCM}=v-u=\frac{1}{2}P_1 $), in the integrable case. Using the known correspondence between integrable theories and electromagnetism, in this section, we construct a closed-form triple-deformed theory that satisfies the flow equation~\eqref{geometries}.

We consider the following deformation of integrable and duality-invariant theories in $d$ dimensions:
\bb
\frac{\partial \L}{\partial \lambda}= \[\frac{1}{d}\(T^{\mu}{}_\nu T^{\nu}{}_\mu-\frac{1}{d} T^{\mu}{}_\mu T^{\nu}{}_\nu\)\]^{\frac{1}{2\alpha}}.
\ee

For simplicity and to allow an explicit comparison with other approaches, we begin with four-dimensional electromagnetic theories. Using the unique formalism of Ref.~\cite{Babaei-Aghbolagh:2025uoz}, we then reproduce the analogous results for two-dimensional integrable theories.  Consequently, for any electromagnetic theory, the duality invariant is given by:
\bb
\frac{1}{4}\(T^{\mu}{}_\nu T^{\nu}{}_\mu-\frac{1}{4} T^{\mu}{}_\mu T^{\nu}{}_\nu\)=\(V \,{\cal L}_V-U\, {\cal L}_U\)^2.
\ee
The equality above can be proven explicitly within the Russo-Townsend framework. In this approach, the operator $\mathcal{R}_\lambda = y\, \Omega^{\prime} (y)$, where $y = e^{\varphi(U,V)}$ and $\varphi(U,V)$ is an auxiliary field,  is introduced, leading to
\begin{equation}\label{lM2}
 y\, \Omega^{\prime} (y)\,=\(V \,{\cal L}_V-U\, {\cal L}_U\) .
\end{equation}
with $\Omega^{\prime} (y)=\frac{1}{y^2} U + V$. An explicit verification of Eq.~\eqref{lM2} then follows from considering ${\cal L}_V=y$ and ${\cal L}_U=- 1/y$. Consequently, we can express the flow equation~\eqref{geometries} as follows:
\bb
\frac{\partial \L}{\partial \lambda}=\pm \Big(V \,{\cal L}_V-U\, {\cal L}_U\Big)^{\frac{1}{\alpha}}. \label{jkj}
\ee
Remarkably, the flow equation in Eq.~\eqref{jkj} can be solved exactly, yielding a closed-form Lagrangian that satisfies an auxiliary equation. The resulting deformed theory and its auxiliary equation are given by:
\bb
\L=-\sqrt{\zeta^2-4 U V}\mp \frac{\alpha-1}{\alpha}\lambda \zeta^{\frac{1}{\alpha}},\label{laggg3}
\ee
where $\zeta$ satisfies the auxiliary equation,
\bb
U-V=\zeta  \sinh \left(\mp \frac{\lambda  \zeta ^{\frac{1}{\alpha }-1}}{\alpha }\right)+\sqrt{\zeta ^2-4 U V} \cosh \left(\mp\frac{\lambda  \zeta ^{\frac{1}{\alpha }-1}}{\alpha }\right)\label{Axee}
\ee
The deformed Lagrangian cannot be expressed in terms of elementary functions. Its expansion to leading orders is given by:
\bb
\L=V-U \mp  \lambda (U + V)^{\frac{1}{\alpha}}+ \frac{\lambda^2}{2 \alpha^2}  \frac{(V-U)  }{(U + V)^{ 2-\frac{2}{\alpha}}}  + \mathcal{O} (\lambda^3) \label{tripla}
\ee
where $\mathcal{O}(\lambda^3)$ denotes higher-order terms.
For the special case $\alpha = 1$, the root-$T\bar{T}$ triple operator in \eqref{geometries} reduces to  root-$T\bar{T}$ flow, and the deformed Lagrangian~\eqref{tripla} recovers the ModMax theory. For $\alpha < 1$, the triple Lagrangian in Eq.~\eqref{tripla} reduces to an irrelevant deformed Lagrangian by setting $\alpha = 1/n$ with $n > 1$. In the even case where $n = 2k$, the resulting theory becomes an analytic theory and invariant under $\varphi$-parity~\cite{Russo:2025fuc} within the Russo-Townsend auxiliary field formalism~\cite{Russo:2024ptw,Babaei-Aghbolagh:2026vkm,Kuzenko:2025gvn,Kuzenko:2026hbw}.
For $\alpha=1/2$, the flow is not the standard Born-Infeld-generating
$T\bar T$ flow~\cite{Conti:2018jho}. Instead, it is governed by the determinant of the traceless stress tensor. Thus, the resulting irrelevant deformation belongs to a distinct $T\bar T$-like class rather than to the ordinary Born-Infeld branch.
 In two dimensions, we obtain a new theory in which the irrelevant flow equation is proportional to the determinant of the traceless energy-momentum tensor, rather than to the determinant of the energy-momentum tensor itself as introduced in Refs.~\cite{Smirnov:2016lqw,Cavaglia:2016oda}.
For $\alpha > 1$, the resulting deformation is relevant and appears here for the first time in the contexts of duality-invariant electromagnetism, $T\bar{T}$ deformations, and integrable theories. No analogous examples have been previously studied within either the Courant-Hilbert approach or the Russo-Townsend auxiliary field formalism for this class of theories.
 Within this auxiliary-field construction, the relevant theory gives rise to a potential $\Omega(\varphi)$ as an explicit function of $\varphi$ expressed as a sum over non-integer powers of $\varphi$:
\begin{equation}
\Omega(\varphi) \sim \sum_{m} a_m \, \varphi^{n/m},
\end{equation}
with $n < m$, and $(n, m) \in \mathbb{Z}$. This constitutes a previously unexplored regime.

Solving the root-$T\bar{T}$ triple flow equation~\eqref{geometries} in two dimensions for $u$ and $v$ yields a two-dimensional integrable deformed Lagrangian analogous to Eq.~\eqref{laggg3} via the transformation $(U, V) \to (u,v)$. 
Substituting the two variables $u$ and $v$ from Eq.~\eqref{uuvv} yields a two-dimensional integrable sigma model in the form of a triple Lagrangian, demonstrating that this Lagrangian describes a unification of the three different types of flow equations:
\bb
\L=\tfrac{1}{2} P_1 + \frac{\lambda^2 P_1 (2 P_2- P_1^2 )^{ \frac{1}{\alpha}-1}}{2^{\frac{2}{\alpha}} \alpha^2} + \frac{\lambda ( 2 P_2- P_1^2 )^{\frac{1}{2 \alpha}}}{2^{\frac{1}{\alpha}}}+ \mathcal{O}(\lambda^3)
\ee

In this section, we unified duality-invariant electromagnetism in four dimensions and integrable theories in two dimensions using a root-$T\bar{T}$ triple flow equation. From Maxwell theory and the principal chiral model, we constructed a closed-form Lagrangian parameterized by $\alpha$ that captures irrelevant, marginal, and relevant deformations. The $\alpha = 1$ case gave $\sqrt{T\bar{T}}$ and ModMax theory; $\alpha < 1$ yielded irrelevant deformations with analyticity and $\varphi$-parity invariance for even denominators; $\alpha > 1$ gave new relevant deformations with non-integer power-law potentials. In two dimensions, the same formalism produced an integrable sigma model that unifies all three classes. 
\section{$\mathcal{R} _\lambda + f(\lambda)\,$ deformations}\label{CCnnn}
Although the Lagrangian is not invariant under duality transformations $\delta \mathcal{L} \neq  0$, it was shown in Ref.~\cite{Gaillard:1981rj} that the derivative of the Lagrangian with respect to a arbitrary duality invariant parameter, such as $(\tau)$, is invariant $\delta( \partial \mathcal{L}/ \partial \tau)=0$. This invariant parameter can be a coupling constant $(\tau \equiv \lambda)$ or an external background field (e.g., the gravitational field $(\tau \equiv g_{\mu \nu})$) that is invariant under duality rotations. Consequently, the energy-momentum tensor, obtained as the derivative of the Lagrangian with respect to the gravitational field, is invariant under duality rotations. 
In the energy-momentum flow approach, the derivative of the Lagrangian with respect to the coupling (a duality-invariant quantity) is expressed in terms of functions of the energy-momentum tensor. 
Because both the energy-momentum tensor and the $\lambda$ coupling are duality-invariant, a general flow equation takes the form
\begin{equation}\label{fgd}
\frac{\partial \mathcal{L}}{\partial \lambda} = \mathcal{O}_\lambda + f(\lambda),
\end{equation}
where $\mathcal{O}_\lambda$ is an operator constructed from the energy-momentum tensor and $f(\lambda)$ is an arbitrary function. Solving this equation explicitly generates theories with generalized flow equations whose Lagrangians obey the duality condition~\eqref{sdc1}. 

 In two dimensions, using the holographic approach, a similar generalization has been made for solvable sigma models.
A solvable generalization of $T\bar{T}$ in two dimensions, known as $T\bar{T} + \Lambda_2$ with $\Lambda_2 = \frac{4(1-\eta)}{\lambda^2}$ where $\eta$ is $(+1)$  for $AdS$ and $\eta$ is $(-1)$  for $dS$, was introduced in Ref.~\cite{Gorbenko:2018oov} and further studied in Refs.~\cite{Lewkowycz:2019xse,Coleman:2021nor,Batra:2024kjl}. This extension of the $T\bar{T}$ flow includes a $\lambda$-dependent cosmological constant, which we implement together with the standard $T\bar{T}$ flow in $\eta=+1$. This generalized deformation is defined for $dS$ background by the flow equation:
\begin{equation}\label{hhjj}
\frac{\partial \mathcal{L}}{\partial \lambda} = \frac{1}{2} \left( T^{\mu}{}_\nu T^{\nu}{}_\mu - T^{\mu}{}_\mu T^{\nu}{}_\nu \right) + \frac{8}{\lambda^2}\,.
\end{equation}

In this section, we present a root-$T\bar{T}$ deformation formulation of the flow equation~\eqref{fgd}, which can be expressed as a ternary deformation parameterized by $\alpha$ as follows:
\begin{equation}\label{geometrieslam}
	\frac{\partial \L}{\partial \lambda}= \Big[\mathcal{R}_\lambda+ f(\lambda)\Big]^{\frac{1}{\alpha}} \equiv \left\lbrace\begin{matrix}
		& irrelevant&\ \ {\rm for\ } \alpha < 1,\cr
		& marginal&\ \ {\rm for\ } \alpha =  1,\cr
		&relevant&\ \  {\rm for\ } \alpha > 1.
	\end{matrix}\right. 
\end{equation}

The strategy for solving differential equations that satisfy the duality-invariant condition in Eq.~\eqref{eq:lagrange_self_duality} with two completely dependent variables is to choose variables that reduce to those in Eq.~\eqref{sdc1} by introducing $U$ and $V$ as defined in Eq.~\eqref{UV}. Therefore, to solve a flow equation in Eq.~\eqref{geometrieslam}, we must select a pair of variables that reduce the flow equation in Eq.~\eqref{geometrieslam} to its simplest form. For this purpose, we consider the variables $P$ and $U$, writing the Lagrangian as $\mathcal{L}(P, U)$. Under this choice, the duality-invariant condition in Eq.~\eqref{eq:lagrange_self_duality} reduces to $  \mathcal{L}^2_P + 2 \frac{U}{P} \, \mathcal{L}_U  \mathcal{L}_P  = -1$.
 In this setting, the root-$T\bar{T}$ operator takes the form $\mathcal{R}_\lambda=\pm\, U \, \mathcal{L}_U $, leading to the flow equation in Eq.~\eqref{geometrieslam} as
\begin{equation}\label{lflowEQa}
	\frac{\partial \L}{\partial \lambda}= \Big[\pm  U  \L_U + f(\lambda)\Big]^{\frac{1}{\alpha}} . 
\end{equation}
 This equation is considerably simpler, as it depends solely on the derivative of the Lagrangian with respect to $U$.

\subsection{A  Relevant Solution}
We present here a relevant solution that confirms the main findings of Section~\eqref{unif}. Taking Maxwell's theory as the seed theory in duality-invariant electrodynamics, we consider $\alpha = 2$, with a minus sign in \eqref{lflowEQa} and  $f(\lambda) = \frac{\lambda^2}{4}$, a relevant theory corresponding to the deformation~\eqref{lflowEQa} is given by
\bb\label{laerr}
\L=-\sqrt{\zeta^2-P^2}+\frac{1}{4} \lambda  \sqrt{\lambda ^2+4 \zeta },
\ee
where $\zeta$ satisfies the auxiliary equation,
\bb\label{aaxxa}
\lambda  \sqrt{\zeta }+\sqrt{\(\frac{\lambda ^2}{\zeta }+4\)\(\zeta ^2-P^2\)}=2 (\frac{P ^2}{4U }-U),
\ee
which is a sixth-order polynomial equation, and so there is no elementary function solution. However, this equation can be solved perturbatively. Substituting $P^2 = 4 U V $ then yields the duality-invariant electrodynamic Lagrangian within the unification formalism, expressed in terms of the two variables $U$ and $V$ as follows:

\begin{eqnarray}\label{Frenedg}
\L&=&V-U +\lambda \sqrt{U + V}  +\lambda^2 \frac{( V-U) }{8 (U + V)}\\ & +&\lambda^3 \frac{(U^2 + 6 U V + V^2) }{32 (U + V)^{5/2}}+...\nonumber
\end{eqnarray}
The effects of the term $f(\lambda)= \frac{1}{4}\lambda^2$ first appear in $\mathcal{O}(\lambda^3)$ in the Lagrangian.

In two-dimensional integrable theories, the variables that yield an equation analogous to Eq.~\eqref{lflowEQa} are $u = 1/4\big( \sqrt{2P_2 - P_1^2} - P_1 \big)$ and $ t = 1 / \sqrt{8}\big( \sqrt{ P_1^2-P_2} \big),$. This gives the triplet root flow 
analogous to Eq.~\eqref{lflowEQa} in 4D. Solving this equation produces a two-dimensional integrable theory consistent with the unification $\mathcal{L}(U,V) \to \mathcal{L}(u, v)$.

In this section, we present a relevant deformed theory—Lagrangian~\eqref{laerr} with auxiliary equation~\eqref{aaxxa}—that satisfies the flow equation~\eqref{geometrieslam} for $\alpha = 2$ and $f(\lambda) = \lambda^2/4$. A triplet solution follows from the characterization approach discussed in the next section.
\subsection{Characteristic Method}
Electromagnetic symmetries and the integrability condition yield the differential equations in Eqs.~\eqref{sdc1} and~\eqref{sdcuv}. Solving these via flow equations built from energy-momentum tensor structures produces a differential equation coupled to $\lambda$, necessitating new mathematical approaches. One such approach is the method of characteristics, first used to solve the $T\bar{T}$ flow equations in two-dimensional integrable theories~\cite{Bonelli:2018kik, Hou:2022csf} and later applied to duality-invariant electromagnetism~\cite{Chen:2024vho, Chen:2025ndc}. In this section, we present a general triple solution to the flow equation~\eqref{geometrieslam} using this method, including irrelevant, marginal, and relevant flow equations for both electromagnetic and integrable theories.

In the characteristic method, both the variables of the differential equation and the average solution are functions of an additional parameter $(s)$. To solve Eq.~\eqref{lflowEQa}, with a minus signature and arbitrary  function  $f(\lambda)$, using the characteristic method, we impose the following characteristic equations for   flow equation~\eqref{lflowEQa}: 
\bb \label{CME}
&\frac{d U}{d s}=U, \qquad \frac{d p_{U}}{d s}=-p_{U},  \qquad \frac{d p_\lambda}{d s}=f^\prime ,\\
&\frac{d \lambda}{d s}=\alpha (p_\lambda)^{\alpha-1}, \qquad \frac{d \L}{d s}=U p_{U} +\alpha (p_\lambda)^{\alpha}.
\ee
where $p_{U}=\pa_{U} \L, p_\lambda=\pa_\lambda \L$.
The initial condition for solving the characteristic equations~\eqref{CME} is:
\bb
&\lambda(s=0)=\lambda_0,\quad U(s=0)=U_0,\\
&\L(s=0)=-U_0+\frac{P^2}{4 U_0},\\
&p_{U}(s=0)=\frac{\pa \L(s=0)}{\pa U_0}=-1-\frac{P^2}{4 U_0^2},\\
&p_\lambda(s=0)=\(-U_0 \,\,p_{U}(s=0)+f(\lambda_0)\)^{1/\alpha}.
\ee
By simplifying the characteristic equations, we obtain a solution to the Lagrangian and an auxiliary equation for $U_0$. This Lagrangian is given by:
\bb\label{fol}
\L=-U_0+\frac{P^2}{4 U_0} + \int _{\lambda_0 }^{\lambda } d \lambda \frac{\alpha f(\lambda)-(1-\alpha)(U_0+\frac{P^2}{4 U_0})}{\alpha\(f(\lambda)+U_0+\frac{P^2}{4 U_0}\)^{\frac{\alpha-1}{\alpha}}},
\ee
with $U_0$ obeying the auxiliary equation:
\bb\label{fffd}
U=U_0 \exp \left(\int _{\lambda_0 }^{\lambda }\frac{d \lambda}{\alpha\(f(\lambda)+U_0+\frac{P^2}{4 U_0}\)^{\frac{\alpha-1}{\alpha}}}\right).
\ee
By eliminating $U_0$ from Eqs.~\eqref{fol} and~\eqref{fffd} and treating $\lambda_0=0$, we obtain the deformed Lagrangian as an explicit function of the two variables $U$ and $P$.

The general Lagrangian in Eq.~\eqref{fol} generates all theories presented in this Letter, as well as further examples of irrelevant and relevant theories. However, the only marginal theory is ModMax, and it is unique.

\section{Fermions}
Consider a free theory of $N_B$ bosons and $N_F$ Dirac fermions in $d = 2$ dimensions:
\begin{equation}\label{fermion_seed}
\mathcal{L}_{\text{seed}} = \frac{1}{2} \partial_\mu \phi^i \partial^\mu \phi^i 
+ \frac{i}{2} \bar{\psi}^I \gamma^\mu \partial_\mu \psi^I 
- \frac{i}{2} \bigl(\partial_\mu \bar{\psi}^I\bigr) \gamma^\mu \psi^I,
\end{equation}
with $i = 1,\dots, N_B$ and $I = 1,\dots, N_F$.
Define the  symmetric  tensor
\begin{equation}
\label{eq:fermionX1}
    (X_1)_{\mu\nu} = \partial_\mu \phi^i \partial_\nu \phi^i 
     + i \bar{\psi}^I \gamma_{(\mu} \partial_{\nu)} \psi^I 
     - i \big(\partial_{(\mu} \bar{\psi}^I\big) \gamma_{\nu)} \psi^I,
\end{equation}
and its square $(X_2)_{\mu\nu} = (X_1)_{\mu\rho} (X_1)^{\rho}{}_{\nu}$. Their traces,
\begin{equation}
    x_1 = (X_1)_\mu{}^\mu, \qquad x_2 = (X_2)_\mu{}^\mu,
\end{equation}
are the only independent trace invariants appearing in the deformed Lagrangian, which thus takes the form $\mathcal{L}(x_1, x_2)$~\cite{Ferko:2022cix}.
Following the electromagnetic analogy, we introduce:
\begin{equation}\label{fermion_UV}
\mathcal{U}_F = \frac{1}{4}\Bigl(\sqrt{2x_2 - x_1^2} - x_1\Bigr), \qquad
\mathcal{V}_F = \frac{1}{4}\Bigl(\sqrt{2x_2 - x_1^2} + x_1\Bigr).
\end{equation}
The triple flow equation is therefore:
\begin{equation}\label{fermion_flow}
\frac{\partial \mathcal{L}}{\partial \lambda} =\pm \bigl(\mathcal{V}_F \mathcal{L}_{\mathcal{V}_F} - \mathcal{U}_F \mathcal{L}_{\mathcal{U}_F}\bigr)^{1/\alpha}.
\end{equation}
This equation is mathematically identical to the bosonic case. Consequently, the solution obtained in Eq.~\eqref{laggg3} applies directly, with $(U,V) \to (\mathcal{U}_F, \mathcal{V}_F)$:
\begin{equation}\label{fermion_lag}
\boxed{
\mathcal{L} = -\sqrt{\zeta^2 - 4\mathcal{U}_F \mathcal{V}_F} \mp \frac{\alpha-1}{\alpha}\lambda \zeta^{1/\alpha},
}
\end{equation}
where $\zeta$ satisfies:
\begin{eqnarray}
\mathcal{U}_F - \mathcal{V}_F &=& \zeta \sinh\!\left(\mp \frac{\lambda \zeta^{1/\alpha - 1}}{\alpha}\right) \nonumber\\
&+& \sqrt{\zeta^2 - 4\mathcal{U}_F \mathcal{V}_F} \cosh\!\left(\mp \frac{\lambda \zeta^{1/\alpha - 1}}{\alpha}\right).
\end{eqnarray}
For $\alpha = 1$, this recovers the fermionic ModMax theory~\cite{Ferko:2023wyi}. These fermionic theories can also be extended to include a cosmological constant, analogous to the integrable sigma model case discussed in Section~\eqref{CCnnn}.

\section{U(1) duality-invariant conformal higher spins}
A general formalism for duality rotations of bosonic conformal spin-$s$ gauge fields ($s \geq 2$) in a conformally flat four-dimensional spacetime was developed in Refs.~\cite{Kuzenko:2021qcx, Kuzenko:2023ebe,Kuzenko:2026vir}. For $s=1$, this reduces to the $\mathsf{U}(1)$ duality-invariant nonlinear electrodynamics of Section~\ref{unif}.

Consider a conformal spin-$s$ field $h_{\alpha(s)\dot{\alpha}(s)} = h_{(\alpha_1\dots\alpha_s)(\dot{\alpha}_1\dots\dot{\alpha}_s)}$. Its field strength is:
\begin{equation}
\mathcal{C}_{\alpha(2s)} = \nabla_{(\alpha_1}{}^{\;\dot{\beta}_1} \cdots \nabla_{\alpha_s}{}^{\;\dot{\beta}_s} h_{\alpha_{s+1}\dots\alpha_{2s})\dot{\beta}(s)},
\end{equation}
where $\nabla_a$ is the conformally covariant derivative. The conjugate field is $\bar{\mathcal{C}}_{\dot{\alpha}(2s)}$.

The free CHS action is:
\begin{equation}
S^{(s)}_{\text{free}}[\mathcal{C},\bar{\mathcal{C}}] = \frac{(-1)^s}{2} \int d^4x \, e \left( \mathcal{C}^{\alpha(2s)}\mathcal{C}_{\alpha(2s)} + \bar{\mathcal{C}}^{\dot{\alpha}(2s)}\bar{\mathcal{C}}_{\dot{\alpha}(2s)} \right).
\end{equation}

For interacting theories, we assume the Lagrangian depends on $\mathcal{C}$ and $\bar{\mathcal{C}}$ only through the two real invariants:
\begin{equation}
\mathfrak{S} = \frac{(-1)^s}{2}\bigl(\mathcal{C}^2 + \bar{\mathcal{C}}^2\bigr), \qquad
\mathfrak{P} = \frac{i}{2}\bigl(\mathcal{C}^2 - \bar{\mathcal{C}}^2\bigr),
\end{equation}
where $\mathcal{C}^2 \equiv \mathcal{C}^{\alpha(2s)}\mathcal{C}_{\alpha(2s)}$.The self-duality condition for deformed CHS theories reduces to~\cite{Kuzenko:2026vir}:
\begin{equation}\label{chs_selfdual}
\mathfrak{P}\,(\mathcal{L}_{\mathfrak{S}}^2 - \mathcal{L}_{\mathfrak{P}}^2 - 1) = 2\mathfrak{S}\,\mathcal{L}_{\mathfrak{S}}\mathcal{L}_{\mathfrak{P}}.
\end{equation}
This equation is structurally identical to Eq.~\eqref{eq:lagrange_self_duality} for the $s = 1$ case under the identification $(\mathfrak{S}, \mathfrak{P}) \to (S, P)$.  We now define the following combinations, analogous to the electromagnetic variables $U$ and $V$:
\begin{equation}\label{chs_UV}
\mathcal{U}_s = \frac{1}{2}\bigl(\sqrt{\mathfrak{P}^2 + \mathfrak{S}^2} - \mathfrak{S}\bigr), \qquad
\mathcal{V}_s = \frac{1}{2}\bigl(\sqrt{\mathfrak{P}^2 + \mathfrak{S}^2} + \mathfrak{S}\bigr).
\end{equation}
In terms of these variables, the self-duality condition in Eq.~\eqref{chs_selfdual} simplifies to the familiar form 
$
\mathcal{L}_{\mathcal{U}_s} \mathcal{L}_{\mathcal{V}_s} = -1,
$
which is identical to Eq.~\eqref{sdc1}. Therefore, the entire triple flow formalism applies. The deformed CHS Lagrangian is:
\begin{equation}
\boxed{
\mathcal{L}^{(s)} = -\sqrt{\zeta_s^2 - 4\mathcal{U}_s\mathcal{V}_s} \mp \frac{\alpha-1}{\alpha}\lambda \zeta_s^{1/\alpha},
}
\end{equation}
with $\zeta_s$ satisfying the same auxiliary equation as before in \eqref{Axee}, with $(U,V) \to (\mathcal{U}_s,\mathcal{V}_s)$. The expansion yields:
\begin{equation}
\mathcal{L}^{(s)} = \mathcal{V}_s - \mathcal{U}_s \mp \lambda (\mathcal{U}_s+\mathcal{V}_s)^{1/\alpha} + \frac{\lambda^2}{2\alpha^2}\frac{(\mathcal{V}_s-\mathcal{U}_s)}{(\mathcal{U}_s+\mathcal{V}_s)^{2-2/\alpha}} + \mathcal{O}(\lambda^3).
\end{equation}

For $s=1$, $\mathcal{U}_1 = U$, $\mathcal{V}_1 = V$, and we recover the electromagnetic case. For $s \geq 2$, this provides the first examples of duality-invariant, irrelevant ($\alpha<1$), marginal ($\alpha=1$), and relevant ($\alpha>1$) deformations of conformal higher-spin theories.
This demonstrates that the triple flow extends to conformal higher-spin field theories, revealing a universal structure underlying duality-invariant theories.
\section{Discussion and Outlook}
In this Letter, we have established a formalism that connects four-dimensional duality-invariant electrodynamics with two-dimensional integrable theories via a root-$T\bar{T}$ triple flow equation. The central result is a closed-form, parameterized Lagrangian that simultaneously describes irrelevant, marginal, and relevant deformations—three classes previously treated as distinct. The parameter $\alpha$ controls the deformation type: $\alpha = 1$ reproduces the known root-$T\bar{T}$ flow and ModMax theory; $\alpha < 1$ yields irrelevant deformations; and $\alpha > 1$ gives a novel family of relevant deformations characterized by non-integer power-law potentials. To our knowledge, this is the first explicit Lagrangian realization of a relevant flow within the $T\bar{T}$ literature.

We have also extended the framework to include a $\lambda$-dependent cosmological constant, and presented a relevant solution for $\alpha = 2$ using the characteristic method. This demonstrates the robustness of our approach and its applicability beyond the simplest flow equations. 
We then generalized the construction to four‑dimensional CHS fields of arbitrary spin $s$, demonstrating that their self‑duality condition reduces to the same universal structure once expressed in terms of $(\mathcal{U}_s ,\mathcal{V}_s)$.

Several promising avenues call for further study. First, the geometric underpinning of the auxiliary equation generated by the root-$T\bar{T}$ triple flow has yet to be fully understood.   Its structural similarity to the Courant-Hilbert and Russo-Townsend formalisms hints at a possible connection to integrable systems and duality symmetries in higher dimensions. We reserve this classification task for a future study. The non-integer power-law potentials appearing in the relevant regime ($\alpha > 1$) suggest a possible link to non-local or non-perturbative phenomena, which could have applications in holography and string theory, where $T\bar{T}$ deformations have been extensively studied. The characteristic method developed here provides a general solution for all three flow classes; it would be interesting to apply it to other nonlinear differential equations in quantum field theory.
 Second, the triple flow framework can be extended to six-dimensional nonlinear chiral 2-form electrodynamics using the results on 6D $T\bar{T}$ deformations in Ref.~\cite{Ferko:2024zth}. These theories admit a direct analog of the duality-invariance condition, which can be cast into Courant-Hilbert form~\cite{Russo:2025wph}. By dimensional reduction and truncation, a generic 6D chiral 2-form theory reduces to a generic duality-invariant 4D electrodynamics, as shown in Ref.~\cite{Bandos:2020hgy}.
Finally, as shown in Ref.~\cite{Bandos:2021rqy, Russo:2024llm}, causality is equivalent to the convexity of $\mathcal{L}$ in the electric field, guaranteeing a Hamiltonian formulation. For any causal, self-dual nonlinear electrodynamics theory, one can construct the Legendre-dual pair $\{\mathcal{L}, \mathcal{H}\}$~\cite{Russo:2025fuc}.
The root-$T\bar{T}$ flow equations induce a Legendre transformation between flowed Lagrangians and Hamiltonians for two-dimensional integrable sigma models~\cite{Tempo:2022ndz, Banerjee:2026qyc}.
We will discuss how the structure of such a root-$T\bar{T}$ triple deformation extends to a phase-space or Hamiltonian formulation. The Hamiltonian formalism for the generalized Lagrangian in Eq.~\eqref{fol}, which includes a cosmological constant, is an important discussion that should also be investigated for integrable sigma models. We defer this classification to future work.
\begin{acknowledgments}
We are grateful to Dmitri Sorokin, Roberto Tateo, Sergei Kuzenko, Jorge G. Russo, and Shahin Sheikh-Jabbari for their interest in this work and the fruitful discussions that followed. The work of H.B.-A. was conducted as part of the PostDoc Program on {\it Exploring TT-bar Deformations: Quantum Field Theory and Applications}, sponsored by Ningbo University. This research was partly supported by NSFC Grant No.12475053, 12235016, and 12588101.
\end{acknowledgments}

\bibliography{Ref}

\end{document}